\documentclass[
aps,%
12pt,%
final,%
notitlepage,%
oneside,%
onecolumn,%
nobibnotes,%
nofootinbib,
superscriptaddress,%
noshowpacs,%
centertags]%
{revtex4}
\begin{document}
\selectlanguage{english} 
\title{Positron supercritical resonances and spontaneous positron creation in slow collisions of heavy nuclei}
\author{\firstname{Dmitry A.} \surname{Telnov} (https://orcid.org/0000-0002-2509-2904)}
\email[]{d.telnov@spbu.ru}
\affiliation{St.~Petersburg State University, St.~Peterburg 199034, Russia}
\author{\firstname{N. K.} \surname{Dulaev} (https://orcid.org/0009-0000-0854-422X)}
\affiliation{St.~Petersburg State University, St.~Peterburg 199034, Russia}
\affiliation{Petersburg Nuclear Physics Institute named by B.~P.~Konstantinov of National Research Center ``Kurchatov Institute'', Gatchina, Leningrad region 188300, Russia}
%
\author{\firstname{Y. S.} \surname{Kozhedub} (https://orcid.org/0000-0003-1273-9008)}
\affiliation{St.~Petersburg State University, St.~Peterburg 199034, Russia}
%
\author{\firstname{I. A.} \surname{Maltsev} (https://orcid.org/0000-0001-8452-639X)}
\affiliation{St.~Petersburg State University, St.~Peterburg 199034, Russia}
%
\author{\firstname{R. V.} \surname{Popov} (https://orcid.org/0000-0003-0730-5213)}
\affiliation{St.~Petersburg State University, St.~Peterburg 199034, Russia}
\affiliation{Petersburg Nuclear Physics Institute named by B.~P.~Konstantinov of National Research Center ``Kurchatov Institute'', Gatchina, Leningrad region 188300, Russia}
%
\author{\firstname{I. I.} \surname{Tupitsyn} (https://orcid.org/0000-0001-9237-5667)}
\affiliation{St.~Petersburg State University, St.~Peterburg 199034, Russia}
%
\author{\firstname{V. M.} \surname{Shabaev} (https://orcid.org/0000-0002-2769-6891)}
\email[]{v.shabaev@spbu.ru}
\affiliation{St.~Petersburg State University, St.~Peterburg 199034, Russia}
\affiliation{Petersburg Nuclear Physics Institute named by B.~P.~Konstantinov of National Research Center ``Kurchatov Institute'', Gatchina, Leningrad region 188300, Russia}
%
\begin{abstract}
We present a theoretical and computational study of positron supercritical resonances in systems consisting of two highly-charged bare nuclei. The resonance positions and widths depending on the internuclear separation are calculated with the help of the complex-scaling generalized pseudospectral method in modified prolate spheroidal coordinates. The results are applied to estimate the probability of spontaneous positron creation in slow U$^{92+}$--U$^{92+}$ and Cm$^{96+}$--Cm$^{96+}$ collisions.
\end{abstract}
\maketitle
\section{Introduction}
The standard quantum electrodynamics (QED) theory predicts that spontaneous electron-positron pair creation by a time-independent electric field can be observed if the field strength is comparable with some critical value or exceeds it \cite{frad1991}. However, to date the spontaneous pair creation has not been detected in experiments due to extremely high value of the critical field strength. For the uniform electric field, it is equal to $1.3\times 10^{18}$~V/m. In the case of the Coulomb field, the supercritical regime can be  achieved in low-energy collisions of two bare nuclei with the total charge number larger than $173$. In such a collision  the originally neutral vacuum should decay into the charged vacuum and two positrons (see Refs.~\cite{pome1945,gers1969,piep1969,zeld1971,grei1985,vosk2021} and references therein). This phenomenon was predicted more than $50$ years ago, but so far it was not observed in experiments. The main obstacle for the experimental observation of this phenomenon is the fact that the time of the supercritical regime is too short. As a result,  the spontaneous pair creation is strongly masked by the dynamical (or induced) pair creation, which is due to the time dependence of the field.  This was  the main reason why the Frankfurt group, which worked on this topic for more than $20$ years, concluded that the vacuum decay can be observed only in collisions with nuclear sticking when the nuclei are bound for some period of time due to nuclear forces \cite{rein2005}. To date, we have no evidence of the nuclear sticking in collisions of so heavy ions, therefore this scenario does not seem promising.  However, in the recent works by the St.~Petersburg group \cite{malt2019,shab2019,popo2020,popo2023} it  was found  that the vacuum decay can be observed via impact-sensitive measurements of pair-production probabilities and corresponding positron spectra. In these papers it was shown that clear signatures indicating the transition from the subcritical to the supercritical regime arise in both pair-creation probabilities and positron spectra if the collisions are considered along the trajectories, which correspond to a given minimum internuclear distance. 

The calculations by the Frankfurt and St.~Petersburg groups showed that the dynamical and spontaneous pair-creation channels contribute coherently, and, generally,  the total pair-creation probability for a given trajectory does not exhibit any threshold effects when crossing the border from the subcritical to the supercritical regime.  For this reason, we cannot formally separate these channels in the total pair-creation probability when considering collisions at given impact parameters and energies. However, to estimate the spontaneous contribution to the total probability, one can integrate the resonance width over the supercritical regime while the nuclei are moving along their trajectories \cite{popo1973,peit1973}. In the present paper, we perform this calculation for the U$^{92+}$--U$^{92+}$ and Cm$^{96+}$--Cm$^{96+}$ head-on  collisions  at  the energies which correspond to the shortest internuclear distance $R_{\rm min} =17.5$~fm. This distance was chosen as the minimal one in the scenarios for the  experimental observation of the transition to the supercritical regime in Refs.~\cite{popo2020,popo2023}. Thus, the present work  should  allow one to better understand  why it is possible to experimentally observe the transition to the supercritical regime  according to the previously proposed  scenarios \cite{malt2019,shab2019,popo2020,popo2023}.

The paper is organized as follows. Section~\ref{th} contains a short account of the theoretical approach, in section~\ref{num} we describe our numerical method, in section~\ref{res} we present and discuss our results, and section~\ref{con} contains concluding remarks. Atomic units are used throughout the paper unless specified otherwise.

\section{Theory}\label{th}
\subsection{Time-independent Dirac equation}
To simplify the calculations, we will use Dirac's hole picture started with the Dirac equation for positrons \cite{grei1985,godu2017}. In this approach, the lower Dirac continuum states as well as discrete states, detached from the lower continuum to the gap between the lower and upper continua, are considered occupied by positrons, which are not free and cannot be observed. A positron becomes free if a transition is made from such states to the upper continuum. At the same time, a hole in the lower continuum or in the discrete state is created, which is described as an electron in a continuum or bound state, respectively. Thus the electron-positron pair creation can be caused by the energy absorption from an external field or may occur spontaneously if the parameters of the nuclear system, such as internuclear separation, make it possible that the energy of some initially occupied positron state get in the upper continuum.

In our case, as the distance between the two nuclei decreases, the energies of the discrete positron states in the gap between the lower and upper continua move towards the onset of the upper continuum. If the total charge of the system $2Z_{n}$ exceeds $173$, then at least the highest-lying $1s\sigma_{g}$ states eventually cross the boundary of the upper continuum, becoming a supercritical resonance and opening up a possibility of spontaneous electron-positron pair creation. The positions and widths of the positron supercritical resonances are obtained by solving the time-independent Dirac equation for a one-positron system with two identical nuclei in the center-of-mass frame of reference:
 \begin{equation}
  H\Psi(\bm{r})=E\Psi(\bm{r}), 
\label{eq10}
 \end{equation}
where $\Psi(\bm{r})$ is a four-component positron wave function and the Hamiltonian $H$ takes the form:
\begin{equation}
 H = c(\bm{\alpha}\cdot\bm{p}) + c^{2}\beta + U. \label{eq20}
\end{equation}
In Eq.~(\ref{eq20}), $c$ is the speed of light (equal to the inverse fine-structure constant in atomic units), $\bm{p}$ is the momentum operator, and $\bm{\alpha}$ and $\beta$ are the Dirac matrices:
\begin{equation}
 \bm{\alpha} = 
 \begin{Vmatrix}
 0_{2} & \bm{\sigma}\\
 \bm{\sigma}& 0_{2}
 \end{Vmatrix},\quad 
 \beta = 
 \begin{Vmatrix}
 1_{2} & 0_{2}\\
 0_{2} & -1_{2}
 \end{Vmatrix}.\label{eq30}
\end{equation}
In Eq.~(\ref{eq30}), $0_{2}$ and $1_{2}$ are zero and unit $2\times 2$ matrices:
\begin{equation}
 0_{2}=\begin{pmatrix}
    0&0\\
    0&0
   \end{pmatrix},\quad
 1_{2}=\begin{pmatrix}
    1&0\\
    0&1
   \end{pmatrix},\label{eq40}
\end{equation}
and $\bm{\sigma}$ denotes the vector consisting of the Pauli matrices as 
components:
\begin{equation}
  \sigma_{x}=\begin{pmatrix}
    0&1\\
    1&0
   \end{pmatrix},
 \ \sigma_{y}=\begin{pmatrix}
    0&-i\\
    i&0
   \end{pmatrix},
 \ \sigma_{z}=\begin{pmatrix}
    1&0\\
    0&-1
   \end{pmatrix}.\label{eq50}
\end{equation}
The interaction with the nuclei $U$ can be described as follows:
\begin{equation}
 U =  U_{n}(|\bm{r}+a\bm{e}_{z}|) +  U_{n}(|\bm{r}-a\bm{e}_{z}|). \label{eq60}
\end{equation}
where $U_{n}$ is the potential which depends on the nuclear charge distribution, $a$ is a half internuclear distance, and $\bm{e}_{z}$ is the unit vector along the $z$ axis.

When solving Eq.~(\ref{eq10}), we are interested in complex energy eigenvalues of the positron resonances:
\begin{equation}
 E=E_{r} - i\frac{\Gamma}{2},
\label{gps55}
\end{equation}
where $E_{r}$ represents the position of the resonance on the energy scale and $\Gamma$ is the resonance width, a positive-definite number, which has a meaning of the positron creation probability per unit time. In theory, such eigenvalues can be obtained by imposing special (outgoing-wave) boundary conditions on the corresponding wave functions at infinity in the coordinate space. However, it appears that the absolute value of the resonance wave function diverges in the real coordinate space as $r\rightarrow\infty$ thus making the numerical implementation of the eigenvalue problem not straightforward since the standard routines assume zero boundary conditions at infinity.  Various approaches exist to solve the problem, such as using complex absorbing potentials \cite{muga2004} and complex scaling of the coordinates \cite{mois1998}. We make use of the uniform complex scaling, when the entire $\lambda$ semiaxis is rotated in the complex plane by some appropriate angle $\Theta$. Upon such rotation, the resonance wave function vanishes at infinity, so conventional zero boundary conditions can be imposed.

\subsection{Nuclear charge distribution model}
For calculations of the supercritical positron resonances by complex scaling, the nuclear potential should allow for analytical continuation. We make use of the Gaussian distribution of the nuclear charge density $\rho_{n} (r)$:
\begin{equation}
 \rho_{n} (r) =\frac{Z_{n}}{\pi^{3/2}}\left(\frac{s}{R_{n}}\right)^{3}\exp\left[-\left(\frac{s r}{R_{n}}\right)^{2}\right],
\end{equation}
where $Z_{n}$ is the total charge of a single nucleus, $R_{n}$ is the root-mean-square nuclear radius, and $s$ is a dimensionless adjustable parameter. In the calculations, we used $R_{n}=5.8571$~fm for the uranium isotope $^{238}$U and $R_{n}=5.8429$~fm for the curium isotope $^{244}$Cm  \cite{ange2013}. In this model, the nuclear potential takes the form:
\begin{equation}
 U_{n}(r)=\frac{Z_{n}}{r}\mathrm{erf} \left(\frac{s r}{R_{n}}\right) ,
\end{equation}
where $\mathrm{erf}(z)$ is the error function. Since the error function is holomorphic on the whole complex plane, its analytical continuation from the real axis is straightforward. The parameter $s$ was adjusted to reproduce the low-lying energy levels of one-electron diatomic quasimolecules at the internuclear separation $2/Z_{n}$ most closely to those given by the Fermi nuclear model \cite{parp1992}, which is considered accurate and is widely used in the calculations \cite{viss1997,shab2013,miro2015}. The best fit is obtained at $s=1.189$. The ground-state ($1s\sigma_{g}$) and the first excited-state ($2p_{1/2}\sigma_{u}$) energies of U$_{2}^{183+}$ and Cm$_{2}^{191+}$ quasimolecules at $R=2/Z_{n}$ are listed in Table \ref{tab1} for the Fermi model and the Gaussian model with $s=1.189$. The energies were computed with the CODATA recommended inverse fine-structure constant value $137.035999084$ \cite{ties2021}.

\section{Numerical method}\label{num}
The eigenvalue problem (\ref{eq10}) is solved with the help of the generalized pseudospectral (GPS) method in prolate spheroidal coordinates \cite{teln2018}. Prolate spheroidal coordinates \cite{abra1972} are a natural choice for description of diatomic quantum systems. Here we make use of slightly modified prolate spheroidal coordinates, more convenient to use for systems with variable internuclear separation. Namely, we shift the traditional coordinate $\xi$ and scale it with the parameter $a$, which is a half internuclear distance:
\begin{equation}
 \lambda = a(\xi -1).
\label{eq130}
\end{equation}
It is assumed that the nuclear charge distributions are centered at the points $\pm a$ on the Cartesian $z$ axis, and the relations between the Cartesian coordinates and modified prolate spheroidal coordinates read as
\begin{equation}
\begin{split}
x&=\sqrt {[(\lambda + a)^{2}-a^{2}][1-\eta^{2}]}\cos\varphi,\\
y&=\sqrt {[(\lambda + a)^{2}-a^{2}][1-\eta^{2}]}\sin\varphi,\\
z&=(\lambda + a)\eta. 
\label{eq140}
\end{split} 
\end{equation}
At any $a$, the coordinates $\lambda$ and $\eta$ vary within the limits $[0,\infty]$ and $[-1,1]$, respectively. The angle $\varphi$, as usually, describes rotation about the $z$ (internuclear) axis and varies within the range $[0,2\pi]$. 

The problem is two-dimensional because the $\varphi$ angle can be eliminated due to rotational symmetry. In the GPS method, the wave functions and operators are discretized on a set of collocation points, which are the roots of orthogonal polynomials or their derivatives. We make use of the Legendre polynomials and apply the Radau quadrature scheme for the $\lambda$ coordinate and Gauss quadrature scheme for the $\eta$ coordinate \cite{teln2009}. Since the collocation points lie within the interval $[-1,1]$, a mapping transformation should be applied to represent the $\lambda$ coordinate:
\begin{equation}
 \lambda = R_{l}\dfrac{(1+x)^{2}}{1-x}\exp(i\Theta).
\label{gps50}
\end{equation}
In Eq.~(\ref{gps50}), the variable $x$ takes the values in the interval $[-1,1]$ and $R_{l}$ is a mapping parameter; the latter is used to change the distribution of the $\lambda$ grid points and thus improve the accuracy of the calculations. The quadratic dependence of $\lambda$ on $1+x$ ensures a dense enough distribution of the grid points at small $\lambda$; in particular, the nuclear potential at very small distances within the nuclear radius can be well described. Note the complex scaling factor $\exp(i\Theta)$ on the right-hand side of (\ref{gps50}). This factor implements the uniform complex scaling in the GPS method, so the complex eigenvalues can be obtained by standard computer routines for diagonalization of non-Hermitian matrices. The complex-scaling GPS method proves to be reliable in calculations of the resonance states in atomic and molecular systems \cite{chus2004}, and the resulting complex energy eigenvalues do not depend on $\Theta$ in a wide range of the complex rotation angle.

The coordinate $\eta$ varies in the same interval $[-1,1]$ where the collocation points of the Gauss--Legendre quadrature lie. It may appear that no mapping transformation is required to represent this coordinate. However, it was pointed out \cite{teln2007} that the wave functions, corresponding to odd orbital angular momentum projections onto the internuclear axis, exhibit a square-root singularity as $\eta\rightarrow\pm 1$, which may significantly reduce the accuracy of numerical differentiation in the GPS method. To circumvent the problem, a special coordinate mapping was suggested \cite{teln2009}, which can be used for any angular momentum projection without loss of accuracy. Here we also apply this mapping for the coordinate $\eta$:
\begin{equation}
 \eta = \sin\left(\frac{\pi}{2}y\right),
\label{gps60}
\end{equation}
where $y$ denotes the collocation points of the Gauss-Legendre quadrature.

We performed calculations of the positron supercritical resonances with the following numerical parameters: $160$ grid points for the coordinate $\lambda$, $16$ grid points for the coordinate $\eta$, the mapping parameter $R_{l}=1/Z_{n}$, the complex rotation angle $\Theta=0.3$.

\section{Results and discussion}\label{res}
As the internuclear distance decreases, the discrete positron vacuum states in the gap between the lower and upper continua move up. The highest-lying $1s\sigma_{g}$ states eventually plunge into the upper continuum, if the total charge of the two nuclei exceeds $173$. If the total charge of the nuclei is even larger, the next discrete positron vacuum states may also cross into the upper continuum. For the collisional systems under consideration with the shortest internuclear distance $R_{\mathrm{min}}=17.5$~fm, this happens for the Cm$^{96+}$--Cm$^{96+}$ system: the $1s\sigma_{g}$ energy level crosses into the upper continuum at the internuclear distance $R\approx 52$~fm, while the $2p_{1/2}\sigma_{u}$ reaches the onset of the upper continuum at $R\approx 20$~fm. For the U$^{92+}$--U$^{92+}$ system, only the $1s\sigma_{g}$ energy level reaches the upper continuum at $R\approx 35$~fm, the $2p_{1/2}\sigma_{u}$ level would do so at $R\approx 11$~fm, but this distance is unreachable since it is smaller than $R_{\mathrm{min}}$.

Once a discrete positron vacuum state crosses into the upper continuum, it becomes an unstable resonance state, which can be characterized by a complex energy (\ref{gps55}). Therefore the vacuum may undergo a spontaneous decay with creation of a free positron. The instantaneous rate of this process is given by $\Gamma$, twice the absolute value of the imaginary part of the complex energy, while the positron energy distribution is centered at $E_{r}$, the real part of the complex energy, and the width of this distribution is equal to $\Gamma$.

\begin{figure}
\setcaptionmargin{5mm} \onelinecaptionsfalse
\captionstyle{normal}
\includegraphics[clip,width=\columnwidth]{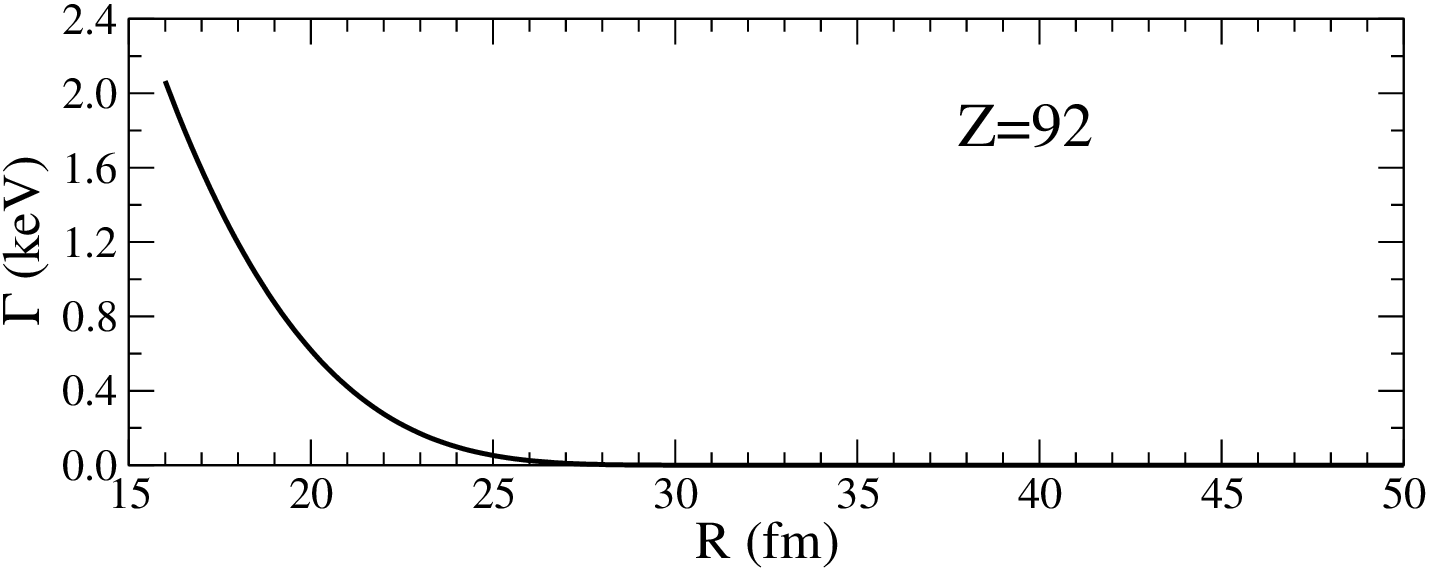}
\includegraphics[clip,width=\columnwidth]{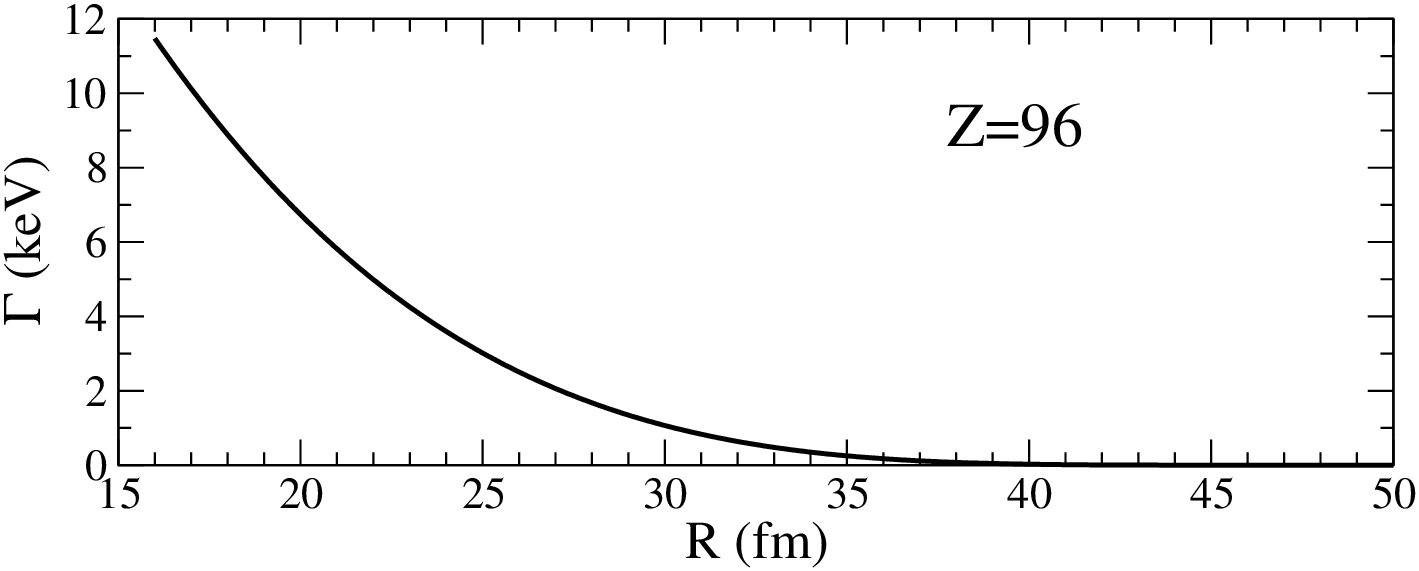}
\caption{Width of $1s\sigma_{g}$ positron supercritical resonance vs internuclear distance in U$^{92+}$--U$^{92+}$ and Cm$^{96+}$--Cm$^{96+}$ systems.}
\label{fig1}
\end{figure}

In Fig.~\ref{fig1}, we show the width of the $1s\sigma_{g}$ positron supercritical resonance as a function of the internuclear distance $R$ for the systems U$^{92+}$--U$^{92+}$ and Cm$^{96+}$--Cm$^{96+}$. For U$^{92+}$--U$^{92+}$, the width $\Gamma$ is identically equal to zero at $R>35$~fm, and for Cm$^{96+}$--Cm$^{96+}$, it is equal to zero at $R>52$~fm. As one can see in Fig.~\ref{fig1}, the resonance width remains quite small unless the resonance energy level plunges deeply into the upper continuum. That is why the contribution of the  $2p_{1/2}\sigma_{u}$ state to the spontaneous positron creation in Cm$^{96+}$--Cm$^{96+}$ can be neglected: this state becomes a supercritical resonance on a small range of internuclear distances $17.5<R<20$~fm, does not go far in the upper continuum, and its width is always several orders of magnitude smaller than that of the $1s\sigma_{g}$ state. Our data for the supercritical $1s\sigma_{g}$ resonance width in the U$^{92+}$--U$^{92+}$ system are in fair agreement with the recent results of Maltsev \textit{et al.} \cite{malt2020}, who used a one-center approach in spherical coordinates and a uniformly charged ball for the nuclear charge distribution, and with the earlier work by Marsman and Horbatsch \cite{mars2011}. Unlike Ref.~\cite{malt2020}, we do not need to search for the optimal complex rotation angle since our results do not depend on this angle within the range $0.2$ to $0.5$.

The spontaneous vacuum decay in slow collisions of heavy nuclei follows the exponential law with the instantaneous rate $\Gamma (t)$ at each time moment $t$ in the supercritical regime, when $\Gamma (t)$ does not vanish. If the probability $P_{s}$ of spontaneous positron creation is well below unity, it can be calculated according to the following equation:
\begin{equation}
 P_{s} =2\int\limits_{-t_{s}}^{t_{s}}dt\ \Gamma (t),
 \label{res10}
\end{equation}
where $-t_{s}$ and $t_{s}$ denote the times of entering and leaving the supercritical regime while the nuclei move along their trajectories. The factor $2$ on the right-hand side of Eq.~(\ref{res10}) accounts for two possible projections of the positron angular momentum on the internuclear axis: the $1s\sigma_{g}$ energy level is degenerate and corresponds to two different states, with the angular momentum projections $\frac{1}{2}$ and $-\frac{1}{2}$ on the internuclear axis. The dependence of $\Gamma$ on time is determined by the law of motion of the nuclei. The function $\Gamma (R)$ presented in Fig.~\ref{fig1} is universal but the law of motion $R(t)$ is different for different trajectories. Below we consider head-on collisions in the U$^{92+}$--U$^{92+}$ and Cm$^{96+}$--Cm$^{96+}$ systems. We assume the classical motion of the nuclei, so their trajectories and laws of motion are the solutions of the well-known Rutherford scattering problem. 

\begin{figure}
\setcaptionmargin{5mm} \onelinecaptionsfalse
\captionstyle{normal}
\includegraphics[clip,width=\columnwidth]{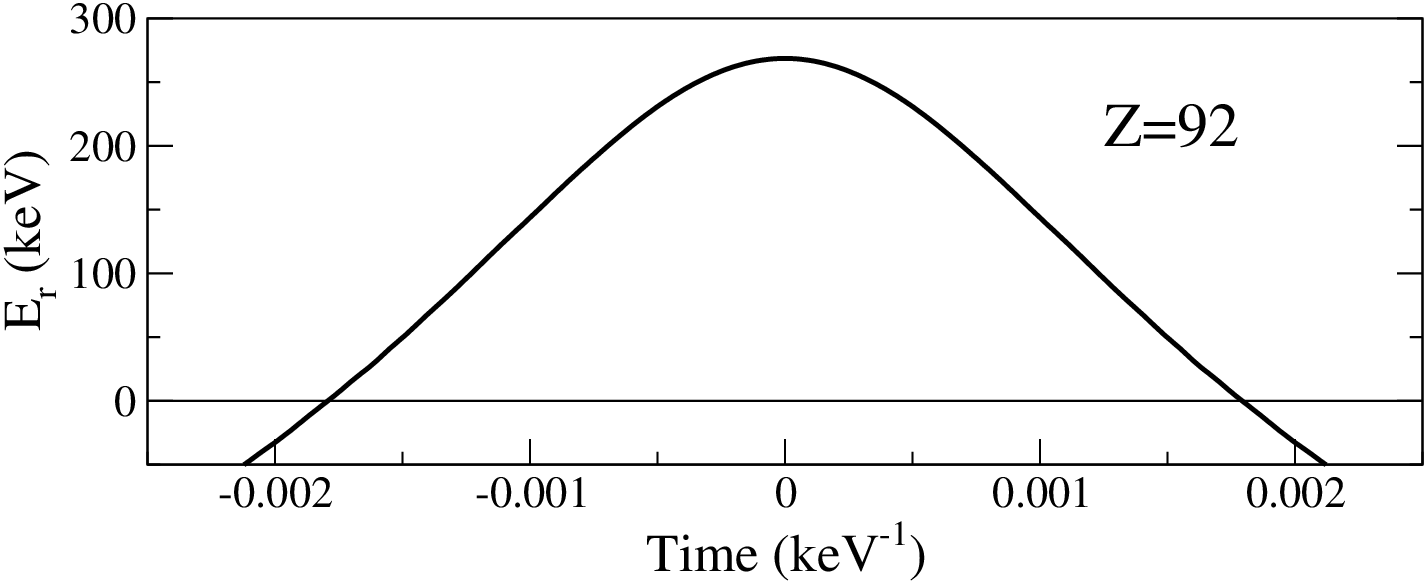}
\includegraphics[clip,width=\columnwidth]{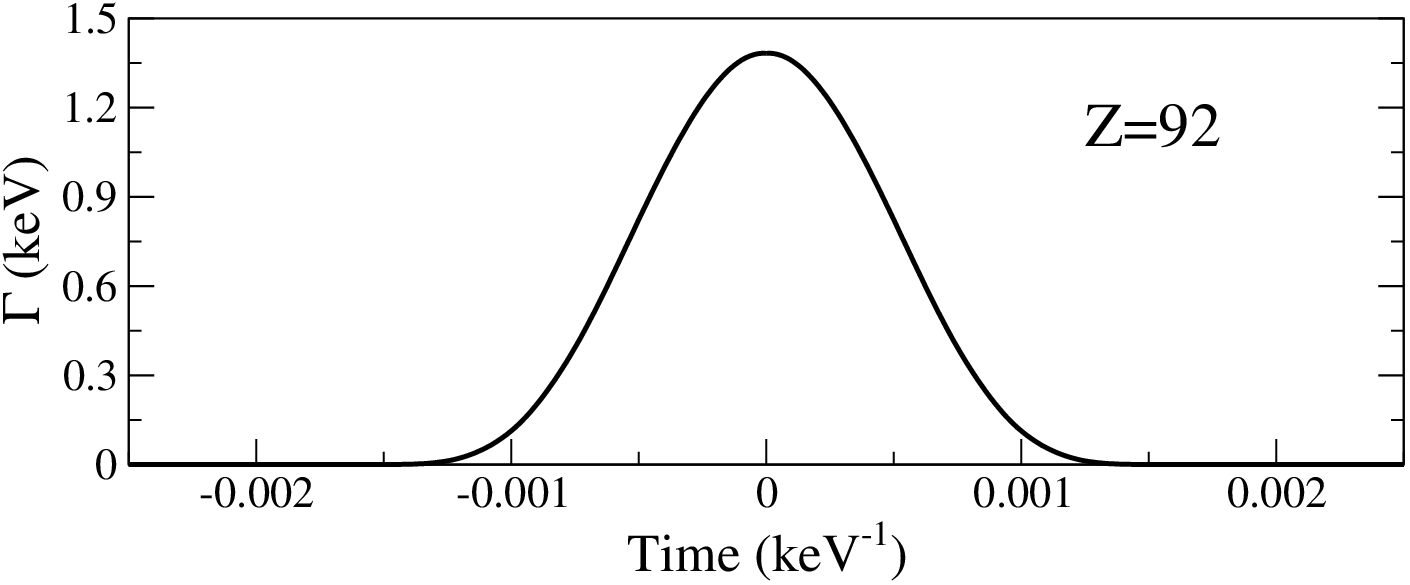}
\caption{Position (upper panel) and width (lower panel) of $1s\sigma_{g}$ positron supercritical resonance vs time for U$^{92+}$--U$^{92+}$ head-on collision with shortest internuclear distance $R_{\mathrm{min}}=17.5$~fm.}
\label{fig2}
\end{figure}

\begin{figure}
\setcaptionmargin{5mm} \onelinecaptionsfalse
\captionstyle{normal}
\includegraphics[clip,width=\columnwidth]{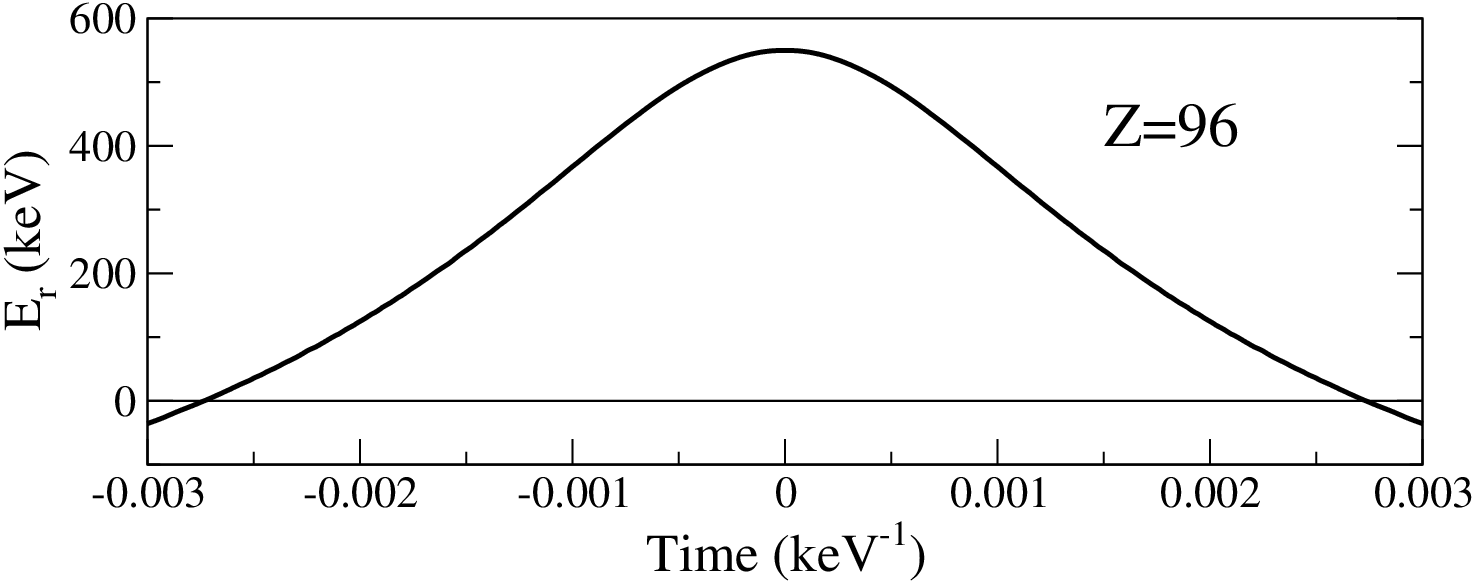}
\includegraphics[clip,width=\columnwidth]{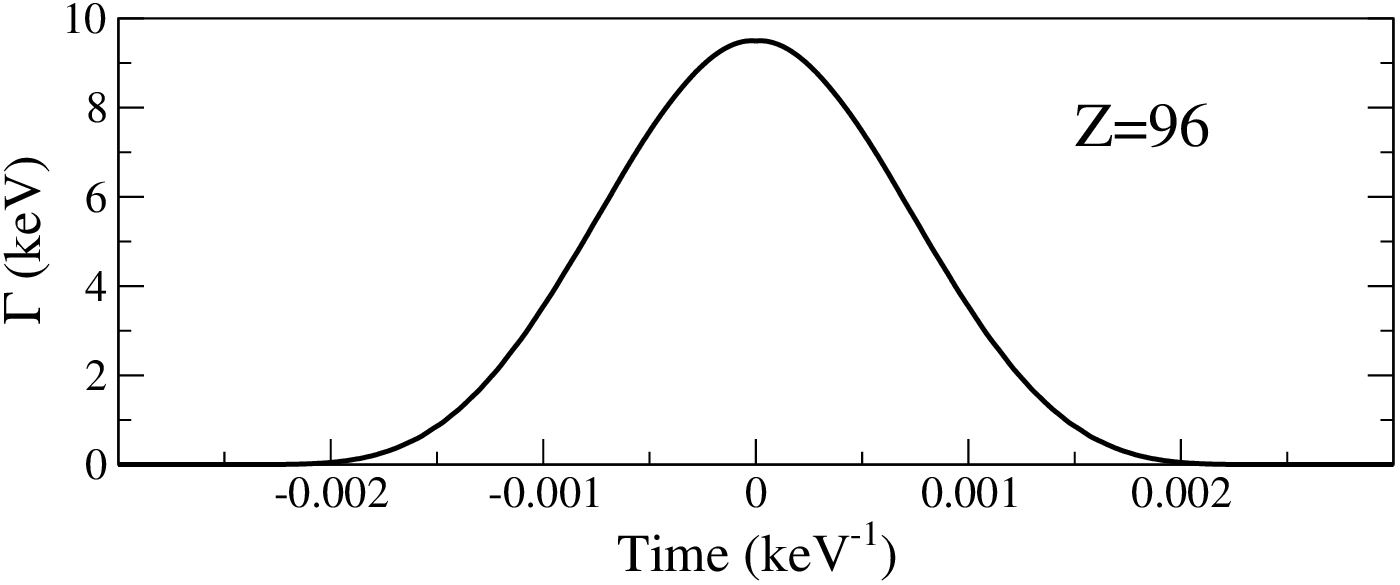}
\caption{Position (upper panel) and width (lower panel) of $1s\sigma_{g}$ positron supercritical resonance vs time for Cm$^{96+}$--Cm$^{96+}$ head-on collision with shortest internuclear distance $R_{\mathrm{min}}=17.5$~fm.}
\label{fig3}
\end{figure}

In Figs.~\ref{fig2} and \ref{fig3}, we show the position of the supercritical resonance $E_{r}$ and its width $\Gamma$ as functions of time for head-on collisions with  $R_{\mathrm{min}}=17.5$~fm in systems U$^{92+}$--U$^{92+}$ and Cm$^{96+}$--Cm$^{96+}$, respectively. The shortest internuclear distance $R_{\mathrm{min}}$ is reached at time $t=0$. For the resonance position $E_{r}$, the zero energy corresponds to the onset of the upper positron continuum. The maximum at $t=0$ in the dependence $E_{r}(t)$ can serve as an estimate of the largest energy of positrons created by the spontaneous mechanism in the collisions under consideration. For the U$^{92+}$--U$^{92+}$ system, this energy is approximately $270$~keV; for the Cm$^{96+}$--Cm$^{96+}$ system, it is equal to $550$~keV. At larger energies, the energy distribution of positrons resulting from the spontaneous vacuum decay is expected to decrease rapidly. 

It is instructive to compare spontaneous positron creation probabilities calculated with the formula (\ref{res10}) and total positron creation probabilities from solving the time-dependent Dirac equation, which can be found for the same head-on collisions in the U$^{92+}$--U$^{92+}$ and Cm$^{96+}$--Cm$^{96+}$ systems in recent papers \cite{popo2020,popo2023}. We show such a comparison in Table~\ref{tab2}. As one can see, the contribution of the spontaneous mechanism to the total positron production is already considerable in the U$^{92+}$--U$^{92+}$ system (about 27\%) and becomes dominant in the Cm$^{96+}$--Cm$^{96+}$ system (about 78\%). These results are fully consistent with the previous findings \cite{malt2019,shab2019,popo2020,popo2023}, which indicate  that the observation of spontaneous vacuum decay is possible with the implementation of some impact-sensitive schemes.

\section{Conclusion}\label{con}
In this paper, we have reported the results of theoretical and computational study of positron supercritical resonances in systems of two bare nuclei. The theoretical approach is based on the time-independent Dirac equation for positrons, and computations are done with the help of the complex-scaling generalized pseudospectral method in modified prolate spheroidal coordinates. The generalized pseudospectral method previously proved to be accurate and efficient in numerous atomic and molecular calculations. The uniform complex scaling in modified prolate spheroidal coordinates implemented in this work appears reliable: the complex energy eigenvalues are insensitive to large variations of the complex rotation angle, at least in the range $0.2$ to $0.5$.

Numerical data, obtained in our calculations of the positron supercritical resonances, are applied to estimate the probabilities of spontaneous positron creation in slow U$^{92+}$--U$^{92+}$ and Cm$^{96+}$--Cm$^{96+}$ collisions. Our results show that contribution of the spontaneous mechanism to the total positron production in head-on collisions with the shortest internuclear separation $R_{\mathrm{min}}=17.5$~fm is significant in the U$^{92+}$--U$^{92+}$ collision and becomes dominant in the Cm$^{96+}$--Cm$^{96+}$ collision. Of course, for nonzero impact parameters, with increasing $R_{\mathrm{min}}$, the dynamic mechanism of positron production becomes more and more important. However, we believe that experimental observation of the fundamental process of spontaneous vacuum decay in slow collisions of highly charged atomic nuclei is still possible if some coincidence scheme is implemented and only positrons resulting from collisions with small impact parameters (and large scattering angles of the nuclei) are collected and analyzed. The transition from the subcritical to supercritical regime can be observed along the scenarios proposed in Refs.~\cite{malt2019,shab2019,popo2020,popo2023}.

\begin{acknowledgments}
 This work was supported by the Russian Science Foundation (Grant No. 22-62-00004).
\end{acknowledgments}

\clearpage
\clearpage
\clearpage
\clearpage
\begin{table}
\setcaptionmargin{0mm} \onelinecaptionsfalse
\captionstyle{flushleft}
\caption{$1s\sigma_{g}$ and $2p_{1/2}\sigma_{u}$ electron energies of U$_{2}^{183+}$ and Cm$_{2}^{191+}$ quasimolecules at $R=2/Z_{n}$  (a.u.). Zero energy is at the onset of the electron upper continuum.}
\bigskip
\begin{tabular}{|c|c|c|c|}
\hline
\multicolumn{1}{|c|}{System}&\multicolumn{1}{c|}{Orbital}&\multicolumn{1}{c|}{Fermi model} &\multicolumn{1}{c|}{Gaussian model}\\
\hline
&&&\\
U$_{2}^{183+}$&$1s\sigma_{g}$&$-9956.962$&$-9956.948$\\
&$2p_{1/2}\sigma_{u}$&$-7187.069$&$-7187.046$\\
\hline
&&&\\
Cm$_{2}^{191+}$&$1s\sigma_{g}$&$-10916.92$&$-10916.93$\\
&$2p_{1/2}\sigma_{u}$&$-8000.568$&$-8000.576$\\
\hline
\end{tabular}
\label{tab1}
\end{table}
\clearpage
\begin{table}
\setcaptionmargin{0mm} \onelinecaptionsfalse
\captionstyle{flushleft}
\caption{Spontaneous positron creation probability $P_{s}$ due to supercritical resonance $1s\sigma_{g}$ and total positron creation probability $P_{t}$ \cite{popo2023} for head-on collisions in U$^{92+}$--U$^{92+}$ and Cm$^{96+}$--Cm$^{96+}$ systems with $R_{\mathrm{min}}=17.5$~fm.}
\bigskip
\begin{tabular}{|c|c|c|c|}
\hline
System&\multicolumn{1}{c|}{$P_{s}$} &\multicolumn{1}{c|}{$P_{t}$}&\multicolumn{1}{c|}{$P_{s}/P_{t}$}\\
\hline
&&&\\
U$^{92+}$--U$^{92+}$&$0.00326$&$0.0120$&$0.27$\\
\hline
&&&\\
Cm$^{96+}$--Cm$^{96+}$&$0.0333$&$0.0425$&$0.78$\\
\hline
\end{tabular}
\label{tab2}
\end{table}


\clearpage
\begin{thebibliography}{30}
\expandafter\ifx\csname natexlab\endcsname\relax\def\natexlab#1{#1}\fi
\expandafter\ifx\csname bibnamefont\endcsname\relax
  \def\bibnamefont#1{#1}\fi
\expandafter\ifx\csname bibfnamefont\endcsname\relax
  \def\bibfnamefont#1{#1}\fi
\expandafter\ifx\csname citenamefont\endcsname\relax
  \def\citenamefont#1{#1}\fi
\expandafter\ifx\csname url\endcsname\relax
  \def\url#1{\texttt{#1}}\fi
\expandafter\ifx\csname urlprefix\endcsname\relax\def\urlprefix{URL }\fi
\providecommand{\bibinfo}[2]{#2}
\providecommand{\eprint}[2][]{\url{#2}}

\bibitem[{\citenamefont{Fradkin \emph{et~al.}}(1991)\citenamefont{Fradkin,
  Gitman, and Shvartsman}}]{frad1991}

\refitem{book}
\bibinfo{author}{\bibfnamefont{E.~S.} \bibnamefont{Fradkin}},
  \bibinfo{author}{\bibfnamefont{D.~M.} \bibnamefont{Gitman}},
  \bibnamefont{and} \bibinfo{author}{\bibfnamefont{S.~M.}
  \bibnamefont{Shvartsman}}, \emph{\bibinfo{title}{Quantum Electrodynamics with
  Unstable Vacuum}} (\bibinfo{publisher}{Springer-Verlag},
  \bibinfo{address}{Berlin}, \bibinfo{year}{1991}).

\bibitem[{\citenamefont{Pomeranchuk and Smorodinsky}(1945)}]{pome1945}

\refitem{article}
\bibinfo{author}{\bibfnamefont{I.}~\bibnamefont{Pomeranchuk}} \bibnamefont{and}
  \bibinfo{author}{\bibfnamefont{J.}~\bibnamefont{Smorodinsky}},
  \bibinfo{journal}{J. Phys. USSR} \textbf{\bibinfo{volume}{9}},
  \bibinfo{pages}{97} (\bibinfo{year}{1945}).

\bibitem[{\citenamefont{Gershtein and Zeldovich}(1969)}]{gers1969}

\refitem{article}
\bibinfo{author}{\bibfnamefont{S.~S.} \bibnamefont{Gershtein}}
  \bibnamefont{and} \bibinfo{author}{\bibfnamefont{Y.~B.}
  \bibnamefont{Zeldovich}}, \bibinfo{journal}{Zh. Eksp. Teor. Fiz.}
  \textbf{\bibinfo{volume}{57}}, \bibinfo{pages}{654} (\bibinfo{year}{1969}),
  \bibinfo{note}{[Sov. Phys. JETP \textbf{30}, 358 (1970)]}.

\bibitem[{\citenamefont{Pieper and Greiner}(1969)}]{piep1969}

\refitem{article}
\bibinfo{author}{\bibfnamefont{W.}~\bibnamefont{Pieper}} \bibnamefont{and}
  \bibinfo{author}{\bibfnamefont{W.}~\bibnamefont{Greiner}},
  \bibinfo{journal}{Z. Phys.} \textbf{\bibinfo{volume}{218}},
  \bibinfo{pages}{327} (\bibinfo{year}{1969}).

\bibitem[{\citenamefont{Zeldovich and Popov}(1971)}]{zeld1971}

\refitem{article}
\bibinfo{author}{\bibfnamefont{Y.~B.} \bibnamefont{Zeldovich}}
  \bibnamefont{and} \bibinfo{author}{\bibfnamefont{V.~S.} \bibnamefont{Popov}},
  \bibinfo{journal}{Usp. Fiz. Nauk} \textbf{\bibinfo{volume}{105}},
  \bibinfo{pages}{403} (\bibinfo{year}{1971}), \bibinfo{note}{[Sov. Phys.
  Uspekhi \textbf{14}, 673 (1972)]}.

\bibitem[{\citenamefont{Greiner \emph{et~al.}}(1985)\citenamefont{Greiner,
  M{\"u}ller, and Rafelski}}]{grei1985}

\refitem{book}
\bibinfo{author}{\bibfnamefont{W.}~\bibnamefont{Greiner}},
  \bibinfo{author}{\bibfnamefont{B.}~\bibnamefont{M{\"u}ller}},
  \bibnamefont{and} \bibinfo{author}{\bibfnamefont{J.}~\bibnamefont{Rafelski}},
  \emph{\bibinfo{title}{Quantum electrodynamics of strong fields}}
  (\bibinfo{publisher}{Springer-Verlag}, \bibinfo{address}{Berlin},
  \bibinfo{year}{1985}).

\bibitem[{\citenamefont{Voskresensky}(2021)}]{vosk2021}

\refitem{article}
\bibinfo{author}{\bibfnamefont{D.~N.} \bibnamefont{Voskresensky}},
  \bibinfo{journal}{Universe} \textbf{\bibinfo{volume}{7}},
  \bibinfo{pages}{104} (\bibinfo{year}{2021}).

\bibitem[{\citenamefont{Reinhardt and Greiner}(2005)}]{rein2005}

\refitem{incollection}
\bibinfo{author}{\bibfnamefont{J.}~\bibnamefont{Reinhardt}} \bibnamefont{and}
  \bibinfo{author}{\bibfnamefont{W.}~\bibnamefont{Greiner}}, in
  \emph{\bibinfo{booktitle}{Proceeding of the Memorial Symposium for Gerhard
  Soff}}, edited by \bibinfo{editor}{\bibfnamefont{W.}~\bibnamefont{Greiner}}
  \bibnamefont{and} \bibinfo{editor}{\bibfnamefont{J.}~\bibnamefont{Reinhardt}}
  (\bibinfo{publisher}{EP Systema}, \bibinfo{address}{Budapest},
  \bibinfo{year}{2005}), pp. \bibinfo{pages}{181--192}.

\bibitem[{\citenamefont{Maltsev \emph{et~al.}}(2019)\citenamefont{Maltsev,
  Shabaev, Popov, Kozhedub, Plunien, Ma, St{\"o}hlker, and Tumakov}}]{malt2019}

\refitem{article}
\bibinfo{author}{\bibfnamefont{I.~A.} \bibnamefont{Maltsev}},
  \bibinfo{author}{\bibfnamefont{V.~M.} \bibnamefont{Shabaev}},
  \bibinfo{author}{\bibfnamefont{R.~V.} \bibnamefont{Popov}},
  \bibinfo{author}{\bibfnamefont{Y.~S.} \bibnamefont{Kozhedub}},
  \bibinfo{author}{\bibfnamefont{G.}~\bibnamefont{Plunien}},
  \bibinfo{author}{\bibfnamefont{X.}~\bibnamefont{Ma}},
  \bibinfo{author}{\bibfnamefont{T.}~\bibnamefont{St{\"o}hlker}},
  \bibnamefont{and} \bibinfo{author}{\bibfnamefont{D.~A.}
  \bibnamefont{Tumakov}}, \bibinfo{journal}{Phys. Rev. Lett.}
  \textbf{\bibinfo{volume}{123}}, \bibinfo{pages}{113401}
  (\bibinfo{year}{2019}).

\bibitem[{\citenamefont{Shabaev \emph{et~al.}}(2019)\citenamefont{Shabaev,
  Bondarev, Glazov, Kozhedub, Maltsev, Malyshev, Popov, Tumakov, and
  Tupitsyn}}]{shab2019}

\refitem{inproceedings}
\bibinfo{author}{\bibfnamefont{V.~M.} \bibnamefont{Shabaev}},
  \bibinfo{author}{\bibfnamefont{A.~I.} \bibnamefont{Bondarev}},
  \bibinfo{author}{\bibfnamefont{D.~A.} \bibnamefont{Glazov}},
  \bibinfo{author}{\bibfnamefont{Y.~S.} \bibnamefont{Kozhedub}},
  \bibinfo{author}{\bibfnamefont{I.~A.} \bibnamefont{Maltsev}},
  \bibinfo{author}{\bibfnamefont{A.~V.} \bibnamefont{Malyshev}},
  \bibinfo{author}{\bibfnamefont{R.~V.} \bibnamefont{Popov}},
  \bibinfo{author}{\bibfnamefont{D.~A.} \bibnamefont{Tumakov}},
  \bibnamefont{and} \bibinfo{author}{\bibfnamefont{I.~I.}
  \bibnamefont{Tupitsyn}}, in \emph{\bibinfo{booktitle}{PoS FFK2019}}
  (\bibinfo{year}{2019}),
  \bibinfo{note}{\url{https://pos.sissa.it/353/052/pdf}}.

\bibitem[{\citenamefont{Popov \emph{et~al.}}(2020)\citenamefont{Popov, Shabaev,
  Telnov, Tupitsyn, Maltsev, Kozhedub, Bondarev, Kozin, Ma, Plunien
  \emph{et~al.}}}]{popo2020}

\refitem{article}
\bibinfo{author}{\bibfnamefont{R.~V.} \bibnamefont{Popov}},
  \bibinfo{author}{\bibfnamefont{V.~M.} \bibnamefont{Shabaev}},
  \bibinfo{author}{\bibfnamefont{D.~A.} \bibnamefont{Telnov}},
  \bibinfo{author}{\bibfnamefont{I.~I.} \bibnamefont{Tupitsyn}},
  \bibinfo{author}{\bibfnamefont{I.~A.} \bibnamefont{Maltsev}},
  \bibinfo{author}{\bibfnamefont{Y.~S.} \bibnamefont{Kozhedub}},
  \bibinfo{author}{\bibfnamefont{A.~I.} \bibnamefont{Bondarev}},
  \bibinfo{author}{\bibfnamefont{N.~V.} \bibnamefont{Kozin}},
  \bibinfo{author}{\bibfnamefont{X.}~\bibnamefont{Ma}},
  \bibinfo{author}{\bibfnamefont{G.}~\bibnamefont{Plunien}},
  \bibnamefont{\emph{et~al.}}, \bibinfo{journal}{Phys. Rev. D}
  \textbf{\bibinfo{volume}{102}}, \bibinfo{pages}{076005}
  (\bibinfo{year}{2020}).

\bibitem[{\citenamefont{Popov \emph{et~al.}}(2023)\citenamefont{Popov, Shabaev,
  Maltsev, Telnov, Dulaev, and Tumakov}}]{popo2023}

\refitem{article}
\bibinfo{author}{\bibfnamefont{R.~V.} \bibnamefont{Popov}},
  \bibinfo{author}{\bibfnamefont{V.~M.} \bibnamefont{Shabaev}},
  \bibinfo{author}{\bibfnamefont{I.~A.} \bibnamefont{Maltsev}},
  \bibinfo{author}{\bibfnamefont{D.~A.} \bibnamefont{Telnov}},
  \bibinfo{author}{\bibfnamefont{N.~K.} \bibnamefont{Dulaev}},
  \bibnamefont{and} \bibinfo{author}{\bibfnamefont{D.~A.}
  \bibnamefont{Tumakov}}, \bibinfo{journal}{Phys. Rev. D}
  \textbf{\bibinfo{volume}{107}}, \bibinfo{pages}{116014}
  (\bibinfo{year}{2023}).

\bibitem[{\citenamefont{Popov}(1973)}]{popo1973}

\refitem{article}
\bibinfo{author}{\bibfnamefont{V.~S.} \bibnamefont{Popov}},
  \bibinfo{journal}{Zh. Eksp. Teor. Fiz.} \textbf{\bibinfo{volume}{65}},
  \bibinfo{pages}{35} (\bibinfo{year}{1973}), \bibinfo{note}{[Sov. Phys. JETP
  \textbf{38}, 18 (1974)]}.

\bibitem[{\citenamefont{Peitz \emph{et~al.}}(1973)\citenamefont{Peitz,
  M{\"u}ller, Rafelski, and Greiner}}]{peit1973}

\refitem{article}
\bibinfo{author}{\bibfnamefont{H.}~\bibnamefont{Peitz}},
  \bibinfo{author}{\bibfnamefont{B.}~\bibnamefont{M{\"u}ller}},
  \bibinfo{author}{\bibfnamefont{J.}~\bibnamefont{Rafelski}}, \bibnamefont{and}
  \bibinfo{author}{\bibfnamefont{W.}~\bibnamefont{Greiner}},
  \bibinfo{journal}{Lett. Nuovo Cimento} \textbf{\bibinfo{volume}{8}},
  \bibinfo{pages}{37} (\bibinfo{year}{1973}).

\bibitem[{\citenamefont{Godunov \emph{et~al.}}(2017)\citenamefont{Godunov,
  Machet, and Vysotsky}}]{godu2017}

\refitem{article}
\bibinfo{author}{\bibfnamefont{S.~I.} \bibnamefont{Godunov}},
  \bibinfo{author}{\bibfnamefont{B.}~\bibnamefont{Machet}}, \bibnamefont{and}
  \bibinfo{author}{\bibfnamefont{M.~I.} \bibnamefont{Vysotsky}},
  \bibinfo{journal}{Eur. Phys. J. C} \textbf{\bibinfo{volume}{77}},
  \bibinfo{pages}{782} (\bibinfo{year}{2017}).

\bibitem[{\citenamefont{Muga \emph{et~al.}}(2004)\citenamefont{Muga, Palao,
  Navarro, and Egusquiza}}]{muga2004}

\refitem{article}
\bibinfo{author}{\bibfnamefont{J.}~\bibnamefont{Muga}},
  \bibinfo{author}{\bibfnamefont{J.}~\bibnamefont{Palao}},
  \bibinfo{author}{\bibfnamefont{B.}~\bibnamefont{Navarro}}, \bibnamefont{and}
  \bibinfo{author}{\bibfnamefont{I.}~\bibnamefont{Egusquiza}},
  \bibinfo{journal}{Phys. Rep.} \textbf{\bibinfo{volume}{395}},
  \bibinfo{pages}{357} (\bibinfo{year}{2004}).

\bibitem[{\citenamefont{Moiseyev}(1998)}]{mois1998}

\refitem{article}
\bibinfo{author}{\bibfnamefont{N.}~\bibnamefont{Moiseyev}},
  \bibinfo{journal}{Phys. Rep.} \textbf{\bibinfo{volume}{302}},
  \bibinfo{pages}{211} (\bibinfo{year}{1998}).

\bibitem[{\citenamefont{Angeli and Marinova}(2013)}]{ange2013}

\refitem{article}
\bibinfo{author}{\bibfnamefont{I.}~\bibnamefont{Angeli}} \bibnamefont{and}
  \bibinfo{author}{\bibfnamefont{K.~P.} \bibnamefont{Marinova}},
  \bibinfo{journal}{At. Data Nucl. Data Tables} \textbf{\bibinfo{volume}{99}},
  \bibinfo{pages}{69} (\bibinfo{year}{2013}).

\bibitem[{\citenamefont{Parpia and Mohanty}(1992)}]{parp1992}

\refitem{article}
\bibinfo{author}{\bibfnamefont{F.~A.} \bibnamefont{Parpia}} \bibnamefont{and}
  \bibinfo{author}{\bibfnamefont{A.~K.} \bibnamefont{Mohanty}},
  \bibinfo{journal}{Phys. Rev. A} \textbf{\bibinfo{volume}{46}},
  \bibinfo{pages}{3735} (\bibinfo{year}{1992}).

\bibitem[{\citenamefont{Visscher and Dyall}(1997)}]{viss1997}

\refitem{article}
\bibinfo{author}{\bibfnamefont{L.}~\bibnamefont{Visscher}} \bibnamefont{and}
  \bibinfo{author}{\bibfnamefont{K.~G.} \bibnamefont{Dyall}},
  \bibinfo{journal}{At. Data Nucl. Data Tables} \textbf{\bibinfo{volume}{67}},
  \bibinfo{pages}{207} (\bibinfo{year}{1997}).

\bibitem[{\citenamefont{Shabaev \emph{et~al.}}(2013)\citenamefont{Shabaev,
  Tupitsyn, and Yerokhin}}]{shab2013}

\refitem{article}
\bibinfo{author}{\bibfnamefont{V.~M.} \bibnamefont{Shabaev}},
  \bibinfo{author}{\bibfnamefont{I.~I.} \bibnamefont{Tupitsyn}},
  \bibnamefont{and} \bibinfo{author}{\bibfnamefont{V.~A.}
  \bibnamefont{Yerokhin}}, \bibinfo{journal}{Phys. Rev. A}
  \textbf{\bibinfo{volume}{88}}, \bibinfo{pages}{012513}
  (\bibinfo{year}{2013}).

\bibitem[{\citenamefont{Mironova \emph{et~al.}}(2015)\citenamefont{Mironova,
  Tupitsyn, Shabaev, and Plunien}}]{miro2015}

\refitem{article}
\bibinfo{author}{\bibfnamefont{D.~V.} \bibnamefont{Mironova}},
  \bibinfo{author}{\bibfnamefont{I.~I.} \bibnamefont{Tupitsyn}},
  \bibinfo{author}{\bibfnamefont{V.~M.} \bibnamefont{Shabaev}},
  \bibnamefont{and} \bibinfo{author}{\bibfnamefont{G.}~\bibnamefont{Plunien}},
  \bibinfo{journal}{Chem. Phys.} \textbf{\bibinfo{volume}{449}},
  \bibinfo{pages}{10} (\bibinfo{year}{2015}).

\bibitem[{\citenamefont{Tiesinga \emph{et~al.}}(2021)\citenamefont{Tiesinga,
  Mohr, Newell, and Taylor}}]{ties2021}

\refitem{article}
\bibinfo{author}{\bibfnamefont{E.}~\bibnamefont{Tiesinga}},
  \bibinfo{author}{\bibfnamefont{P.~J.} \bibnamefont{Mohr}},
  \bibinfo{author}{\bibfnamefont{D.~B.} \bibnamefont{Newell}},
  \bibnamefont{and} \bibinfo{author}{\bibfnamefont{B.~N.}
  \bibnamefont{Taylor}}, \bibinfo{journal}{Rev. Mod. Phys.}
  \textbf{\bibinfo{volume}{93}}, \bibinfo{pages}{025010}
  (\bibinfo{year}{2021}).

\bibitem[{\citenamefont{Telnov \emph{et~al.}}(2018)\citenamefont{Telnov,
  Krapivin, Heslar, and Chu}}]{teln2018}

\refitem{article}
\bibinfo{author}{\bibfnamefont{D.~A.} \bibnamefont{Telnov}},
  \bibinfo{author}{\bibfnamefont{D.~A.} \bibnamefont{Krapivin}},
  \bibinfo{author}{\bibfnamefont{J.}~\bibnamefont{Heslar}}, \bibnamefont{and}
  \bibinfo{author}{\bibfnamefont{S.-I.} \bibnamefont{Chu}},
  \bibinfo{journal}{J. Phys. Chem. A} \textbf{\bibinfo{volume}{122}},
  \bibinfo{pages}{8026} (\bibinfo{year}{2018}).

\bibitem[{\citenamefont{Abramowitz and Stegun}(1972)}]{abra1972}

\refitem{book}
\bibinfo{editor}{\bibfnamefont{M.}~\bibnamefont{Abramowitz}} \bibnamefont{and}
  \bibinfo{editor}{\bibfnamefont{I.~A.} \bibnamefont{Stegun}}, eds.,
  \emph{\bibinfo{title}{Handbook of {Mathematical} {Functions} with {Formulas},
  {Graphs}, and {Mathematical} {Tables}}} (\bibinfo{publisher}{Dover},
  \bibinfo{address}{New York}, \bibinfo{year}{1972}), \bibinfo{edition}{10th}
  ed.

\bibitem[{\citenamefont{Telnov and Chu}(2009)}]{teln2009}

\refitem{article}
\bibinfo{author}{\bibfnamefont{D.~A.} \bibnamefont{Telnov}} \bibnamefont{and}
  \bibinfo{author}{\bibfnamefont{S.-I.} \bibnamefont{Chu}},
  \bibinfo{journal}{Phys. Rev. A} \textbf{\bibinfo{volume}{80}},
  \bibinfo{pages}{043412} (\bibinfo{year}{2009}).

\bibitem[{\citenamefont{Chu and Telnov}(2004)}]{chus2004}

\refitem{article}
\bibinfo{author}{\bibfnamefont{S.-I.} \bibnamefont{Chu}} \bibnamefont{and}
  \bibinfo{author}{\bibfnamefont{D.~A.} \bibnamefont{Telnov}},
  \bibinfo{journal}{Phys. Rep.} \textbf{\bibinfo{volume}{390}},
  \bibinfo{pages}{1} (\bibinfo{year}{2004}).

\bibitem[{\citenamefont{Telnov and Chu}(2007)}]{teln2007}

\refitem{article}
\bibinfo{author}{\bibfnamefont{D.~A.} \bibnamefont{Telnov}} \bibnamefont{and}
  \bibinfo{author}{\bibfnamefont{S.-I.} \bibnamefont{Chu}},
  \bibinfo{journal}{Phys. Rev. A} \textbf{\bibinfo{volume}{76}},
  \bibinfo{pages}{043412} (\bibinfo{year}{2007}).

\bibitem[{\citenamefont{Maltsev \emph{et~al.}}(2020)\citenamefont{Maltsev,
  Shabaev, Zaytsev, Popov, Kozhedub, and Tumakov}}]{malt2020}

\refitem{article}
\bibinfo{author}{\bibfnamefont{I.~A.} \bibnamefont{Maltsev}},
  \bibinfo{author}{\bibfnamefont{V.~M.} \bibnamefont{Shabaev}},
  \bibinfo{author}{\bibfnamefont{V.~A.} \bibnamefont{Zaytsev}},
  \bibinfo{author}{\bibfnamefont{R.~V.} \bibnamefont{Popov}},
  \bibinfo{author}{\bibfnamefont{Y.~S.} \bibnamefont{Kozhedub}},
  \bibnamefont{and} \bibinfo{author}{\bibfnamefont{D.~A.}
  \bibnamefont{Tumakov}}, \bibinfo{journal}{Opt. Spectrosc.}
  \textbf{\bibinfo{volume}{128}}, \bibinfo{pages}{1100} (\bibinfo{year}{2020}).

\bibitem[{\citenamefont{Marsman and Horbatsch}(2011)}]{mars2011}

\refitem{article}
\bibinfo{author}{\bibfnamefont{A.}~\bibnamefont{Marsman}} \bibnamefont{and}
  \bibinfo{author}{\bibfnamefont{M.}~\bibnamefont{Horbatsch}},
  \bibinfo{journal}{Phys. Rev. A} \textbf{\bibinfo{volume}{84}},
  \bibinfo{pages}{032517} (\bibinfo{year}{2011}).

\end{thebibliography}
\end{document}